\documentclass[twocolumn,superscriptaddress,aps,pre]{revtex4-2}
\usepackage{url}
\usepackage[utf8]{inputenc}
\usepackage[american,british]{babel}
\usepackage[T1]{fontenc}
\usepackage[pdftex]{graphicx}  
\usepackage{graphicx, xcolor}
\usepackage{dcolumn}
\usepackage{physics}
\usepackage{braket}
\usepackage{bm}
\usepackage{amsmath,amsthm,amssymb}
\usepackage{color}
\usepackage{verbatim} 
\usepackage{soul} 

\usepackage{svg}
\usepackage{placeins}  

\usepackage{natbib}

\newcommand{\im}{\rm{i}}

\newcommand{\smt}{Science, Mathematics and Technology Cluster, Singapore University of Technology and Design, 8 Somapah Road, 487372 Singapore}
\newcommand{\epd}{Engineering Product Development Pillar, Singapore University of Technology and Design, 8 Somapah Road, 487372 Singapore}
\newcommand{\cqt}{Centre for Quantum Technologies, National University of Singapore 117543, Singapore}
\newcommand{\majulab}{MajuLab, CNRS-UNS-NUS-NTU International Joint Research Unit, UMI 3654, Singapore} 
    
\definecolor{darkGreen}{RGB}{0,110,0}
\definecolor{darkBlue}{RGB}{0,0,130}

\usepackage{hyperref}
\hypersetup{
 colorlinks=true,
 linkcolor=blue,
 anchorcolor = blue,
 citecolor = blue,
 filecolor = blue,
 urlcolor = blue
}

\usepackage{dsfont}

\makeatletter
\renewcommand{\fnum@figure}{Fig. \thefigure}
\makeatother

\def \be {\begin{equation}} 
\def \ee {\end{equation}}

\usepackage{tabularx}

\newcommand{\aod}{\hat{a}^\dagger}
\newcommand{\aop}{\hat{a}}
\newcommand{\bdw}{\hat{b}^\dagger(\omega)}
\newcommand{\bw}{\hat{b}(\omega)}
\newcommand{\dm}{\hat{\rho}}
\newcommand{\hop}{\hat{H}}

\begin{document}

\title{Emergence of thermodynamic functioning regimes from finite coupling between a quantum thermal machine and a load}

\author{Gauthameshwar S.} 
\affiliation{\smt} 

\author{Noufal Jaseem} 
\affiliation{\smt} 
\affiliation{\cqt}

\author{Dario Poletti}
\email{dario\_poletti@sutd.edu.sg}
\affiliation{\smt} 
\affiliation{\epd} 
\affiliation{\cqt} 
\affiliation{\majulab}

\begin{abstract}  
Autonomous quantum thermal machines are particularly suited to understand how correlations between thermal baths, a load, and a thermal machine affect the overall thermodynamic functioning of the setup. 
Here, we show that by tuning the operating temperatures and the magnitude of the coupling between machine and load, the thermal machine can operate in four modes: engine, accelerator, heater, or refrigerator. 
In particular, we show that as we increase the coupling strength, the engine mode is suppressed, and the refrigerator mode is no longer attainable, leaving the heater as the most pronounced functioning modality, followed by the accelerator. 
This regime switching can be amplified by quantum effects, such as the bosonic enhancement factor for a harmonic oscillator load, which modifies the effective machine-load coupling, making the thermodynamic functioning sensitive to the initial preparation of the load.
\end{abstract}

\maketitle

\begin{figure}[!htb]
    \includegraphics[width=0.77\linewidth]{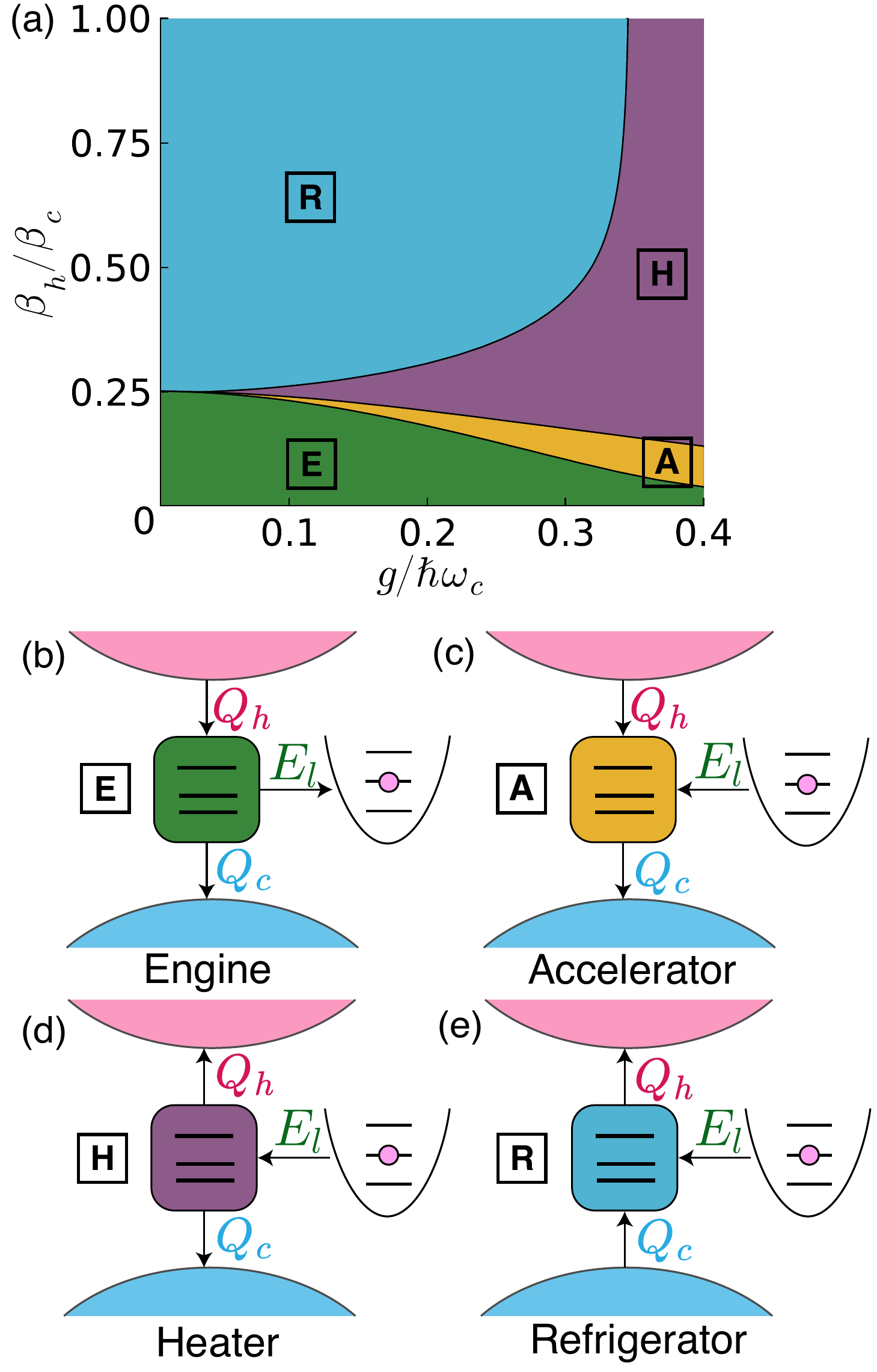}
    \caption{
    (a) Diagram of the functioning regimes of the thermal machine as a function of coupling strength $g$ and inverse temperature of the hot bath $\beta_h$. We observe four functioning regimes from bottom to top: engine (green), accelerator (yellow), heater (purple), refrigerator (blue), respectively labeled by E, A, H, and R and depicted in panels (b-e). Other system parameters are $\eta = 0.005\omega_c$, $\beta_c = 2 (\hbar\omega_c)^{-1}$, and $\omega_l=\omega_e = 3\omega_c$. 
    }
    \label{fig:fig1}
\end{figure}

{\it Introduction}: Since the advent of the Industrial Revolution, thermodynamic machines have significantly influenced our lives. Furthermore, the efforts to understand their functioning have also changed how we think about the world, from the laws of thermodynamics to the concept of entropy \cite{muller2007TDHistory}. 
A typical paradigm for the study of thermodynamics 
brings together a thermal machine, where the working fluid is confined, one hot and one cold bath, which generate the non-equilibrium conditions needed for the machine's functioning, and a load where energy is transferred to or taken from. 
Thermal machines can function in four different regimes: engines, refrigerators, heaters, and accelerators, as shown in Fig.~\ref{fig:fig1}(b,e) and in Refs.~\cite{campisi2020QubitOttoEngine, watson2025quantummachines, menon2025QuantumStatisticsEngine, campisi2023FermionEnginee}.  

With the advent of quantum mechanics, a natural question is how the functioning of thermal machines changes at the quantum scale \cite{binder2018TDReview, Kosloff2013, DeffnerCampbell2019, VinjanampathyAnders2016, Roadmap2025, myers2022QuantumDevices, Cangemi2024QuantumEngsNRefs}, and this has also led to remarkable experimental work \cite{Pekola2014, Klatzow2019, Maslennikov2019, vonLindenfels2019, peterson2019ExperimentalOttoSpin, QuantumLoad2020}. 
Quantum thermal machines have been studied in three main regimes: periodically driven, continuously driven, and autonomous. While periodically driven thermodynamic machines allow for a clear mapping to the ``classical'' thermodynamic cycles, and continuously driven are particularly relevant in the study of thermoelectricity, it is via autonomous engines, for which the load is also modeled exactly, that one can gain insight on how the interaction and correlations between the machine and the load can affect the performance of the thermodynamic machine. 
Preliminary steps in this direction were taken theoretically \cite{colinTeo2017, scarani2017, seah2018QuantumRotorEngines} and experimentally \cite{QuantumLoad2020}. In these works, it was shown that the correlations between engine and load increase the entropy of the load. Consequently, one typically does not only observe a transfer of work, but heat transfers are entwined with work transfers.

Here, we demonstrate that increasing the interaction strength between the thermodynamic machine and the quantum load induces qualitative changes in the machine's operational behavior as illustrated in Fig.~\ref{fig:fig1}(a). 
We also identify a genuinely quantum enhancement of this phenomenology. When the load is a harmonic oscillator, the machine's operation becomes sensitive to the load’s initial occupation, due to bosonic enhancement arising from the action of creation and annihilation operators, which effectively amplifies the coupling strength between the machine and the load. Notably, this effect vanishes when the load is modeled as a ladder system or treated classically.

\begin{figure}[!htb]
    \includegraphics[width=0.95\linewidth]{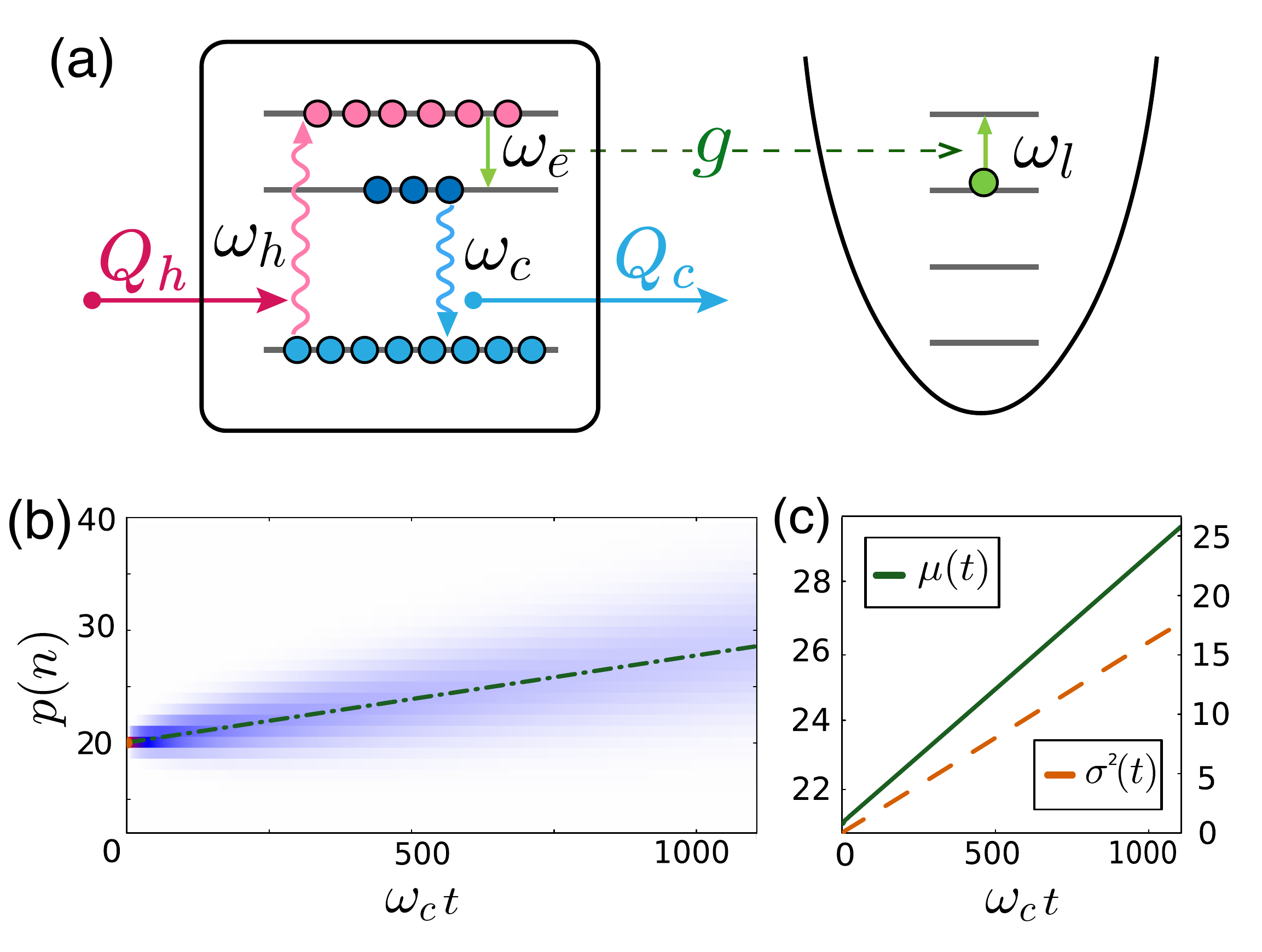}
    \caption{(a) Depiction of a three-level autonomous machine coupled to an oscillator (the load) and exchanging energy with the two baths when functioning as an engine. (b) Typical evolution of the probability of occupation of energy levels $p(n)$ of the load as a function of time, and (c) typical evolution of the mean $\mu(t)$ and variance $\sigma^2(t)$ of the energy in the load, when the thermodynamic machine functions as an engine. Common parameters are $\eta = 0.005\omega_c$, $\beta_h = 0.4(\hbar\omega_c)^{-1}$, $\beta_c = 2 (\hbar\omega_c)^{-1}$, $g=0.05\hbar\omega_c$ and $\omega_l=\omega_e = 3\omega_c$.
    }
    \label{fig:fig2}
\end{figure}

{\it Model}: We study a three-level engine coupled to a quantum load, which, unless stated otherwise, is modeled by a harmonic oscillator, as depicted in Fig.~\ref{fig:fig2}(a).  
The overall Hamiltonian is $\hat{H} = \hat{H}_M + \hat{H}_L  + \hat{H}_{I}$, where, respectively, the thermal machine, load, and their interaction Hamiltonian terms are given by 
$\hat{H}_M = \sum_{i=1}^3 E_i |i\rangle\langle i|$, 
$\hat{H}_L = \hbar \omega_l \left(\hat{a}^\dagger \hat{a} + 1/2\right)$, and 
\begin{align}
\hat{H}_I = g \left(|2\rangle\langle 3| + |3\rangle\langle 2| \right)\left( \aop + \aod \right). \label{eq:engineloadinteraction}    
\end{align}
We set $E_i$ such that $E_1 = 0$, $E_2 = \hbar \omega_c$, $E_3= \hbar \omega_h = \hbar (\omega_c + \omega_e)$, and consequently, $E_3 - E_2 = \hbar \omega_e$. $\aod$ and $\aop$ are the creation and destruction operators for the harmonic oscillator load whose energy spacing is set to $\hbar\omega_l$. 
The engine is coupled to two uncorrelated thermal bosonic baths via the coupling Hamiltonian term $\hop_j$, 
\begin{equation}
    \hop_j =  \hat{A}_{j}\int_0^{\infty} d\omega \; g(\omega) \left[\bdw + \bw\right],  \label{eq:enginebathcoupling}
\end{equation}
for $j \in \{ h, c \}$. The thermal machine's interaction operators with the hot and cold baths are 
\(
\hat{A}_h = \ketbra{1}{3} + \ketbra{3}{1},\)
 and 
\(
\hat{A}_c = \ketbra{1}{2} + \ketbra{2}{1}. 
\)
Each bath contains an uncorrelated and continuous set of positive modes of the oscillators $\omega$ whose creation and destruction operators are \(\bdw\) and \(\bw\). For each bath, we have an Ohmic spectral density with exponential cut-off $S(\omega)= \eta \omega e^{-\omega/\omega_c}$, and the hot and cold baths are set to a Gibbs thermal state with inverse temperatures $\beta_h$ and $\beta_c$, respectively. 
Since the thermal machine and the load have a finite coupling between them, we use a global master equation in Gorini-Kossakowski-Sudarshan-Lindblad form \cite{Gorini1976, Lindblad1976} to model the effects of the baths on the density operator describing the machine and the load $\dm_S$. The dynamics of our system is thus given by \cite{breuerBookOQS, dario2022BoundaryDrivenSystems} 
\begin{equation}
\dot{\dm}_S = \frac{\im}{\hbar}\left[\dm_S, \hop \right] + \sum_{j=h, c} \sum_{\omega} \gamma_{j}(\omega) \mathcal{D}_j[\omega](\dm_S),
\label{eq: GLME in main}
\end{equation}
where 
\(
\mathcal{D}_j[\omega](\dm_S) = 
\hat{A}_j(\omega) \dm_S \hat{A}_j^\dagger(\omega) 
- \frac{1}{2} \{ \hat{A}_j^\dagger(\omega) \hat{A}_j(\omega), \dm_S \},
\) and 
\(
\hat{A}_j(\omega) = \sum_{\epsilon' - \epsilon = \hbar\omega} \bra{\epsilon} \hat{A}_j \ket{\epsilon^{\prime}} \ket{\epsilon}\bra{\epsilon^{\prime}}
\) 
is the jump operator's decomposition in the eigenstates $\ket{\epsilon}$ of the total Hamiltonian of the machine plus the load that corresponds to an energy difference of $\hbar\omega$, and $\gamma_j(\omega)$ is the spectral function associated with absorbing energy $\hbar\omega$ from the $j$-bath, see Apps.~\ref{app:theoryOqs}, \ref{app:bosonic_bath_functions} for more details. 
We evaluate the heat energy drawn into the machine from the respective baths using 
\begin{equation}
    \dot{Q}_j(t) = \sum_\omega \gamma_j(\omega)\Tr\left[\hop \mathcal{D}_j[\omega](\dm_S(t))\right]. 
\end{equation}
To quantify the work exchanged with the load, we evaluate the rate of change of the load's energy. In general, the work that can be extracted from the load does not correspond to its energy, and ergotropy ~\cite{ergotropy2004, ergotropy2020} should be the quantity to be measured. Ergotropy is given by  
\begin{equation}
\mathcal{W}_l(\dm_L) = \Tr(\dm_L \hop_L) - \Tr(\hat{\pi}(\dm_L)\hop_L)
\label{eq:ergotropy-formula}
\end{equation}
where $\hat{\pi}(\dm_L)$ is the passive state obtained by sorting $\dm_L$'s eigenvalues in descending order against $\hop_L$'s eigenenergies. 
However, in our setup, the energy levels occupation of the load follows a biased diffusion dynamics, exemplified in Fig.~\ref{fig:fig2}(b,c), as also seen in \cite{colinTeo2017, QuantumLoad2020}. The average occupation of the load $\mu(t)$ increases (decreases) linearly in time when functioning as an engine (refrigerator/accelerator/heater), while the variance of the occupation $\sigma^2(t)$ also increases linearly for any thermodynamic functioning mode. 
We can thus write the rate of increase of the energy of the load $P_l$ and the rate of change of the ergotropy $\dot{\mathcal{W}}_l$ in terms of drift velocity $v = d\mu(t)/dt$ and diffusion coefficient $D = d\sigma^2(t)/dt$ as, see also App.~\ref{app: classical biased diffusion analysis},  
\begin{eqnarray}
    P_l &=& \frac{d}{dt}\left[\Tr(\hop_L \dm_S)\right] = \hbar\omega_l v \\
    \dot{\mathcal{W}}_l(t) &=& \hbar\omega_l \left(v  - \sqrt{\frac{2D}{\pi t}}\right).     \label{eq: powers}
\end{eqnarray}
Importantly, the rate of change in ergotropy converges, over long times, to that of the energy, and this is why we consider $v$ as a key quantity, together with the heat exchanges $\dot{Q}_j$, to characterize the functioning of the thermal machine. 

\begin{figure}[htp]
    \includegraphics[width=0.8\linewidth]{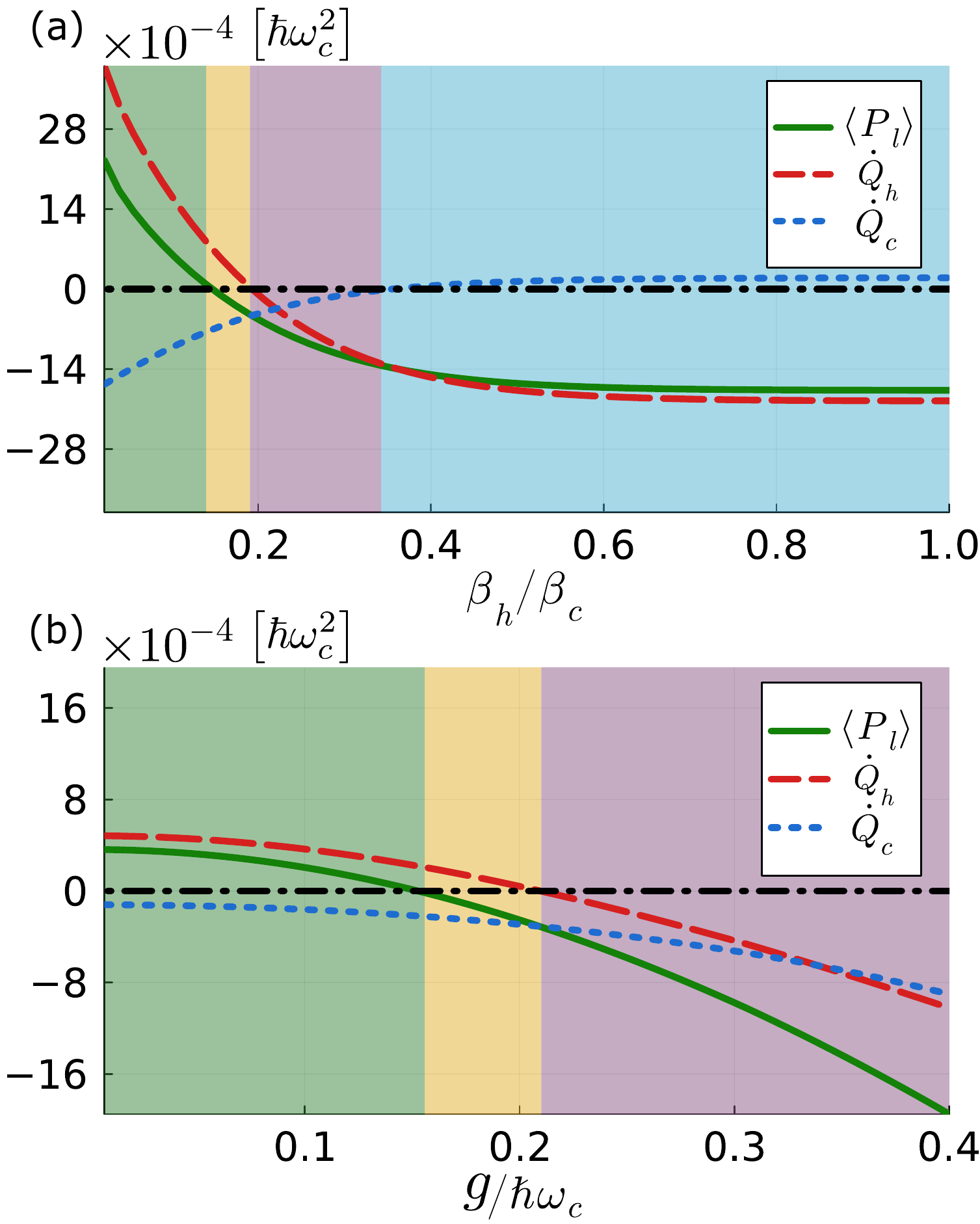}
    \caption{
    Energy currents $\dot{Q}_h$, $\dot{Q}_c$, and the rate of increase of energy in the load $P_l$ as functions of system parameters. (a) At fixed coupling strength $g/\hbar\omega=0.25$, and (b) at fixed inverse temperature ratio $\beta_h/\beta_c = 0.20$.     
    The common parameters are $\eta = 0.005\omega_c$, $\beta_c = 2.0 (\hbar\omega_c)^{-1}$, and $\omega_l=\omega_e = 3.0\omega_c$.} 
    \label{fig:fig3}
\end{figure}

{\it Results}: We keep the energy level spacing of the three-level system fixed to $\omega_e = 3\omega_c$, also $\beta_c=2(\hbar\omega_c)^{-1}$ and perform a parameter sweep over $\beta_h$ and $g$ to map the steady-state of the system's heat currents to the operational regimes. As shown in Fig.~\ref{fig:fig1}(a), at weak coupling $g/\hbar \omega_c \ll 1$, we observe the expected change from engine to refrigerator behavior at the temperature $\beta_h/\beta_c = \omega_c/\omega_h=1/4$~\cite{Scovil1959, kosloff1994ThreeLevelEngine, kosloff1996HeatPump, Geva2002HeatEngines, Varinder2020ThreeLevelEngine}, see also App.~\ref{app:lme}. However, as $g$ increases, the phase space boundary between engine and refrigerator gives rise to the emergence of two intermediate regimes: accelerator and heater. For $g/\hbar \omega_c \lesssim 0.344$, as we increase $\beta_h$, the engine regime changes into a refrigerator through the intermediate regimes of accelerator and heater, as shown in Fig.~\ref{fig:fig3}(a). This occurs as the directions of the energy transfers, $P_l,\;\dot{Q}_h,\;\dot{Q}_c$, sketched in Fig.~\ref{fig:fig1}(b,e), change one by one. For $g/\hbar \omega_c \gtrsim  0.344$, the refrigerator functioning regime disappears for all values of $\beta_h$ and the functioning diagram is dominated by the heater. 
Fig.~\ref{fig:fig3}(b), where $\beta_h/\beta_c$ is kept fixed at $0.2$, also demonstrates how the coupling strength $g$ acts as a control parameter that can be used to switch between thermodynamic functioning regimes. 
This behavior is absent in models employing local GKSL master equations to model the weak coupling between the machine and the load (see App.~\ref{app:lme}).

\begin{figure}[!htp]
    \centering
    \includegraphics[width=0.9\linewidth]{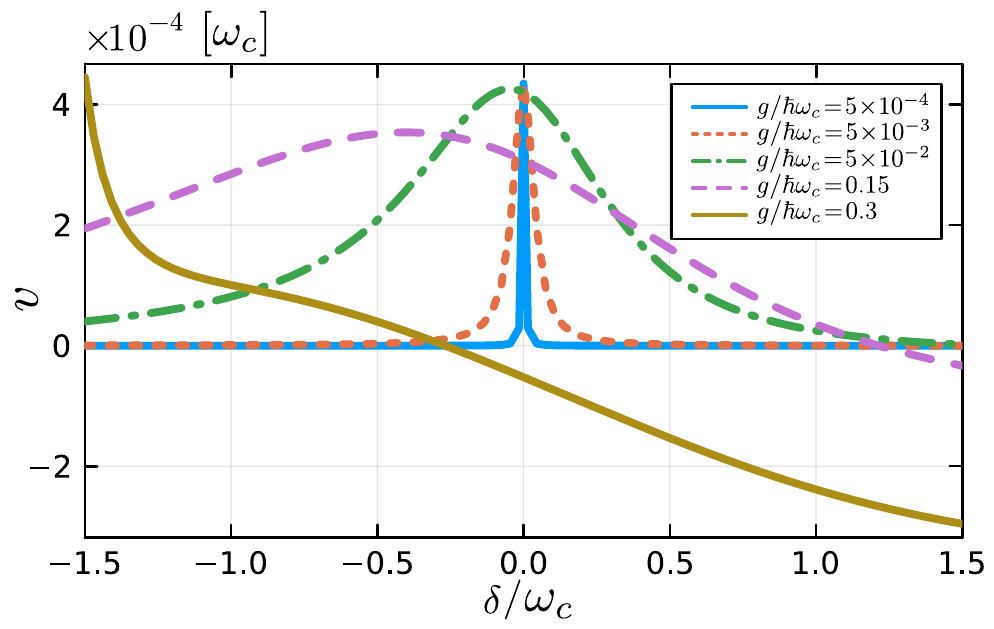}
    \caption{Drift velocity $v$ as functions of the detuning parameter $\delta$. 
    Other system parameters are $\eta = 0.005\omega_c$, $\beta_c = 2 (\hbar\omega_c)^{-1}$, $\beta_h = 0.15\beta_c$ and $\omega_e = 3\omega_c$. 
    }
    \label{fig: detuning}
\end{figure}

The detuning response of the load reveals deeper insights into the crossover between functioning regimes. We rewrite the load frequency $\omega_l$ in terms of a detuning parameter $\delta$ relative to the three-level transition $\omega_e$  
\begin{equation}
    \omega_l = \omega_e + \delta. 
\end{equation}
Fig.~\ref{fig: detuning} shows the drift velocity $v$ as a function of $\delta$ for various $g$. 
In the weak-coupling regime ($g/\hbar\omega_c = 5\times 10^{-4}, 5\times 10^{-3}, 5\times 10^{-2}$), the response follows a Lorentzian profile peaked at resonance $\delta=0$. The width of the Lorentzian increases with $g$, accompanied by a slight redshift of the peak. 
This behavior was analytically shown for a periodically driven three-level system where the driving amplitude played the analogous role of $g$ in our autonomous setup~\cite{kosloff1994ThreeLevelEngine}. However, as the magnitude of the coupling $g$ is increased, the peak progressively becomes asymmetric and flattens ($g/\hbar\omega_c = 0.15$), and eventually ($g/\hbar\omega_c = 0.3$), at $\delta=0$, one can even observe a negative drift velocity. 
The disappearance of the Lorentzian peak profile for larger coupling indicates that the resonant engine mechanism is suppressed and replaced by higher-order processes that drive the load's energy increase rate in the opposite direction ($v<0$). 

The dependence of the functioning on the coupling to the load raises another important question. Since the load is a harmonic oscillator, the coupling between the thermodynamic machine and the load is affected by the load occupation too, via the bosonic enhancement factor $\aop |n\rangle = \sqrt{n}|n-1\rangle$, where $|n\rangle$ is the $n$-th level of the harmonic oscillator. 

\begin{figure}[htp]
    \includegraphics[width=0.90\linewidth]{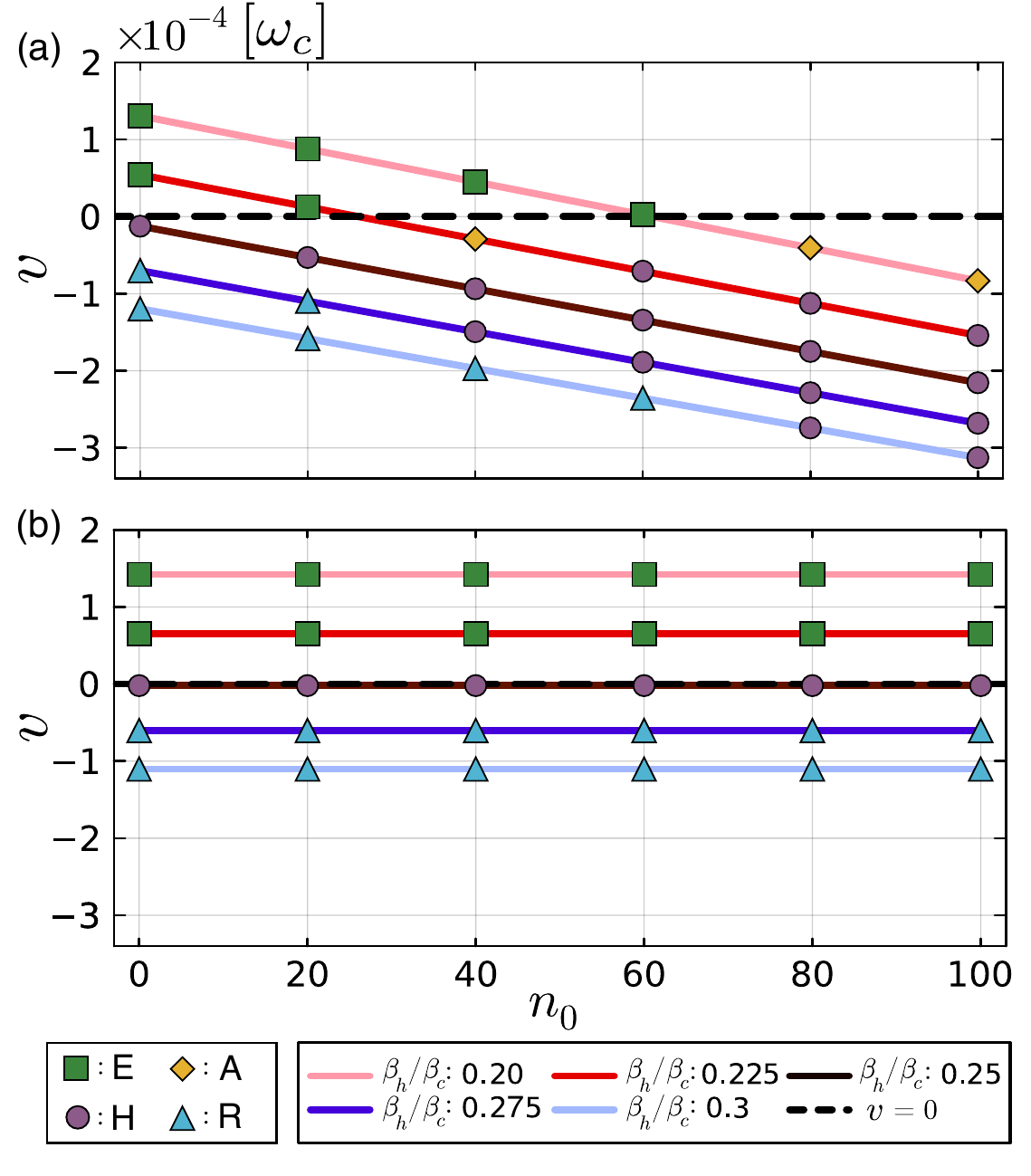}
    \caption{Drift velocity $v$ as a function of initial offset $n_0$. Different lines from top to bottom correspond to different (inverse) temperatures of the hot bath $\beta_h$, from $\beta_h/\beta_c=0.2$ to $0.3$. 
    The symbols correspond to different functioning regimes: green square for engine, yellow diamond for accelerator, purple circle for heater, and blue triangle for refrigerator. 
    Panels (a) and (b) are respectively for a harmonic oscillator or a ladder load. 
    Other system parameters are $\eta = 0.005\omega_c$, $\beta_c = 2 (\hbar\omega_c)^{-1}$, $g = 0.05 \hbar\omega_c$ and $\omega_l=\omega_e = 3\omega_c$. 
    }
    \label{fig: offsetting-the-load}
\end{figure}

To investigate this, we study the dynamics of the setup when the load is initialized at a higher occupation number by modifying the matrix elements of the bosonic operators $\aop$ via $\sqrt{n}\rightarrow \sqrt{n + n_0}$, while ensuring the absence of finite size effects. 
This offset causes an effective increase in the coupling strength which, for large off-settings, is approximated by $g\rightarrow g\sqrt{n_0}$~\footnote{
This relation can be analytically derived by the Taylor expansion of $\sqrt{n_0 + n}$ under the approximation $n_0 \gg n$. We also verified this by comparing two equivalent cases: (i) offsetting the ladder by $n_0$ while keeping $g$ fixed, and (ii) increasing $g$ to $c_0 g\sqrt{n_0}$ without offset. 
}. 
As a result, the functioning regime of the thermodynamic machine changes when varying the offset $n_0$, in the same way as moving horizontally in the phase space shown in Fig.~\ref{fig:fig1}(a). For example, increasing $n_0$ mimics a larger machine-load coupling. 
To verify this, we evaluate the power into the load, quantified by $v$, as a function of $n_0$. Our results are depicted in Fig.~\ref{fig: offsetting-the-load}(a) for different values of the temperature of the hot bath (from $\beta_h/\beta_c=0.2$, top, to $\beta_h/\beta_c=0.3$, bottom). Here we observe that $v$ decreases linearly as $n_0$ increases, going from positive values, where the machine functions as an engine (depicted by green squares), to negative values, where it can function as an accelerator (yellow diamonds), heater (purple circles), or refrigerator (blue triangles).  
Finally, we change the load from a harmonic oscillator to a ladder, whereby the creation and destruction operators $\aod, \aop$ are substituted by raising and lowering operators $\hat{d^{\dagger}}, \hat{d}$, respectively, without the occupation-dependent bosonic enhancement factors in $\hop_I$ such that $\hat{d}\ket{n} = \ket{n-1}$. For this case, the coupling between the machine and the load would not depend on the occupation, and we do not expect any dependence of the functioning regime on the initial occupation. This is depicted in Fig.~\ref{fig: offsetting-the-load}(b), where we observe no effect of the initialization of the load on the functioning of the thermal machine.

{\it Conclusions}: We have studied the thermodynamic functioning of a three-level system coupled to two thermal baths and one load. We have shown that in the presence of finite coupling between the three-level system and the load, one can observe four different thermodynamic functioning regimes. A load that is prepared in a highly non-passive state, when coupled to the baths, would tend to decrease its occupation unless the machine acts as an engine. For larger coupling between the machine and the load, the machine cannot push the load to higher energies, and the load's energy decreases. This results in an accelerator and, for even larger couplings, into a heater whereby the power lost from the load surpasses the magnitude of both the heat currents between the baths, forcing all energy from the load to go into each bath. For a refrigerator, a larger coupling between load and the machine allows energy transfer from the load to the heat bath too, thus making the refrigerator's functioning regime unattainable at larger couplings. 

Interestingly, we observe a clear quantum effect, that the engine's functioning depends on the occupation of the load. In a classical regime where the load is occupied at much higher excitation levels, the bosonic enhancement factor, which induces this effect, becomes negligible. 
Importantly, this dependence on the initial condition does not affect, up to the scales of time and load size we have been able to study, the emergence of a linear increase or decrease of the load's energy. This is because even when the load is driven by the thermal machine to a certain large occupation, the setup has nonzero coherence, unlike when initializing the load at some occupation.         

Since the harmonic oscillator load is bounded from below, as the average occupation decreases, the lowest levels will start to be occupied, and thus the linear decrease will cease. We have thus observed, not shown here, that the overall setup relaxes to a steady state in both the machine and the load. The problem thus becomes about characterizing the relaxation dynamics and studying the properties of a non-equilibrium steady state. 
Furthermore, we would like to stress here that we are studying a setup with {\it finite} coupling between the thermal machine and the load. We prefer not to refer to it as {\it strong} coupling, as in the latter case, the engine and the load cannot be readily distinguished from one another, thus rendering the interpretation of their dynamics through the lens of thermodynamics highly nontrivial. 

As for experimental realization, three-level engines or refrigerators were already considered in the interpretation of masers as thermodynamic machines \cite{scovil1959ThreeLevelRef}. More recently, NV centers were used to realize a three-level engine experimentally \cite{Klatzow2019}, and the magnitude of the coupling to the load could be increased by using a cavity. 
Superconductive three-level systems have also been realized experimentally \cite{Sillanpaa2009, Kumar2016, Xu2016, Vepsalainen2019}. 
Trapped ions \cite{Blatt2012} and optomechanical systems \cite{Paternostro2022} also offer a venue for experimental realization of this setup, possibly using two two-level systems as a thermal machine instead of a single three-level system. 
Despite these advances, fundamental questions remain open. For instance, as the coupling between components of a thermal machine increases, it becomes unclear where the boundary lies between distinct subsystems and a single, inseparable entity. We currently lack effective tools to model, analyze, or interpret such more comprehensive entities. 
Furthermore, while we have discussed clear quantum effects such as the relevance of the bosonic enhancement factor, we also anticipate that finite coupling between the thermal machine and load will influence their thermodynamic functioning at a classical level in nanoscopic systems. Exploring these effects will be part of our future work. 

{\it Acknowledgments}: We thank D. Aghamalyan, K. Bharti, A. Eckardt, V. Scarani, and R. Uzdin for fruitful discussions. D.P. acknowledges the support of the Ministry of Education, Singapore, under the grant MOE-T2EP50123-0017. The simulations were performed using \textit{QuantumOptics.jl}, a Julia-based numerical framework for simulating open quantum systems\cite{kraemer2018quantumoptics.jl}. The computational work for this Letter was performed on the resources of the National Supercomputing Centre, Singapore \cite{nscc}.   


\bibliographystyle{apsrev4-2}
\bibliography{main}


\setcounter{equation}{0}
\setcounter{figure}{0}     
\setcounter{section}{0}
\renewcommand{\thefigure}{A\arabic{figure}}
\renewcommand{\thesection}{\arabic{section}}

\begin{appendix}

\section{Derivation of the Global Master Equation}
\label{app:theoryOqs}

We outline the derivation of the Redfield and GKSL-type master equations used to describe the dynamics of a quantum system weakly coupled to thermal reservoirs. The full derivation, based on standard techniques (see, e.g., \cite{breuerBookOQS}), starts with the system-bath Hamiltonian
\begin{equation}
\hat{H} = \hat{H}_S + \hat{H}_B + \lambda \hat{H}_{SB}, \quad \hat{H}_{SB} = \sum_\alpha \hat{A}_\alpha \otimes \hat{B}_\alpha
\end{equation}
Here, $\lambda$ is a dimensionless parameter capturing the strength of system-bath interaction. Working in the interaction picture of \( \hat{H}_0 = \hat{H}_S + \hat{H}_B \), and assuming factorized initial conditions and stationary bath state due to weak coupling (Born approximation), the reduced dynamics of the system can be expressed to second order in \(\lambda\) as
\begin{equation}
\dot{\dm}_S(t) = -\lambda^2 \int_0^t \mathrm{d}t'\, \mathrm{Tr}_B \left[ \hop_{SB}(t), \left[ \hop_{SB}(t'), \dm_S(t') \otimes \dm_B \right] \right], 
\label{eq: BM_dynamics}
\end{equation}
with $\dm_B\propto e^{-\beta \hat{H}_B}$, and $\beta$ being the inverse temperature of the bath. 
We can collect the traced-out effect of the bath $\text{Tr}_B(\cdot)$ in our system inside the bath correlation function defined by
\begin{equation}
    C_{\alpha\beta}(t - t') = \mathrm{Tr}_B\left\{ \hat{B}_\alpha(t) \hat{B}_\beta(t') \dm_B \right\},
\end{equation}
and simplify Eq.~(\ref{eq: BM_dynamics}) in terms of these functions after invoking the Markov approximation, we shift to a memoryless version of the equation (Redfield-II form \cite{dario2022BoundaryDrivenSystems}):
\begin{equation}
\begin{aligned}
    \dot{\dm}_S(t) = -\lambda^2 \sum_{\alpha\beta} \int_0^\infty &\mathrm{d}\tau \Big(
    C_{\alpha\beta}(\tau) [\hat{A}_\alpha(t), \hat{A}_\beta(t-\tau) \dm_S(t)] \\
    +\;&C_{\beta\alpha}(-\tau) [\dm_S(t) \hat{A}_\beta(t-\tau), \hat{A}_\alpha(t)]
    \Big).
\end{aligned}
\label{eq:Redfield}
\end{equation}

We now express the interaction operators in the eigenbasis of the system Hamiltonian \( \hop_S = \sum_\epsilon \epsilon\, \Pi(\epsilon) \), where \( \Pi(\epsilon) = \ketbra{\epsilon}{\epsilon} \) projects onto the eigenspace of energy \(\epsilon\). Let the interaction operators \(\hat{A}_{\alpha}(t)\) be defined in this eigenbasis via
\begin{equation}
    \hat{A}_\alpha(t) = \sum_\omega e^{i\omega t} \hat{A}_\alpha(\omega), \quad \hat{A}_\alpha(\omega) = \sum_{\epsilon' - \epsilon = \omega} \Pi(\epsilon) \hat{A}_\alpha \Pi(\epsilon').
\end{equation}

Substituting this into Eq.~\eqref{eq:Redfield} and collecting secular (energy-conserving) terms only, we arrive at the global master equation (GME) in GKSL form   
\begin{eqnarray}
\dot{\dm}_S &=& \sum_{\omega,\alpha,\beta} \gamma_{\alpha\beta}(\omega) \left(
    \hat{A}_\beta(\omega) \dm_S \hat{A}_\alpha^\dagger(\omega) \right.\nonumber\\
    &\qquad& \qquad \qquad - \frac{1}{2} \left. \{ \hat{A}_\alpha^\dagger(\omega) \hat{A}_\beta(\omega), \dm_S \}
\right) \nonumber\\
&\equiv& \mathcal{L}(\dm_S),
\label{eq:  GLME}
\end{eqnarray}
where the real part of the correlation function
\begin{align}
    \gamma_{\alpha\beta}(\omega) &= \lambda^2\int_{-\infty}^\infty \mathrm{d}\tau\, e^{i\omega \tau} C_{\alpha\beta}(\tau).
\end{align}

Despite guaranteeing positivity, hermiticity, and unit trace of the density matrices, GME can result in thermodynamic inconsistencies, such as non-conservation of local observables and improper description of system coherences, and the Redfield-II equation is often preferred for its more accurate handling of thermodynamically consistent non-equilibrium state dynamics~\cite{dario2022BoundaryDrivenSystems, tupkary2022LindbladRedfield, chiara2018QME, levy2014LocalLaws}. We have compared the results from the GME with those from Redfield-II, and we have observed quantitative agreement, while the GME calculations can be performed about 40 times faster. This validates our choice of the GME to describe the system we studied. 


\setcounter{figure}{0}  
\renewcommand{\thefigure}{B\arabic{figure}}
\section{The bosonic-bath spectral function}
\label{app:bosonic_bath_functions}

Our engine couples two levels of the three-level system with the respective bosonic baths. We can compute the spectral function of the above interaction term by considering a spin-boson model with an $\sigma^x_{ij}-x$ interaction Hamiltonian         
\[
\hat{H}_{SB,ij} = \hat{\sigma}^x_{ij} \otimes \int_0^{\infty} g(\omega)(\bdw + \bw) \, d\omega.
\]
Here, the $\hat{\sigma}^x_{ij} = \ketbra{i}{j} + \ketbra{j}{i}$ couples two different levels of the three-level system, $i \neq j$. $\bw$ and $\bdw$ are the destruction and creation operators of the bath oscillator with frequency $\omega$. The bosonic modes are uncorrelated at different frequencies and each is initialized in a Gibbs thermal state $\hat{\rho}_B = \hat{G}_{\beta}(\omega) = e^{-\beta \hat{H}_B(\omega)}/Z$ with $Z=\Tr\left(e^{-\beta \hat{H}_B(\omega)}\right)$ and $\hat{H}_B(\omega) = \hbar\omega \left[\bdw \bw+1/2\right]$. Thus, the bath correlation function becomes      
\begin{equation}
\begin{aligned}
    C_{\omega\omega'}(t) = \; & \delta(\omega-\omega') \Tr[\hat{B}(\omega,t)\hat{B}(\omega', 0)\hat{G}_{\beta}(\omega')]\\
    = \; & \delta(\omega - \omega')g(\omega)g(\omega') \nonumber \\ 
    &\times \left(e^{i\omega t} \bar{N}(\omega )+ e^{-i\omega t}(1 + \bar{N}(\omega))\right)
\end{aligned}
\end{equation}
where we used 
    $\hat{b}(\omega, t) = e^{-i\omega t}\hat{b}(\omega)$, and \(\quad
    \langle \hat{b}^{\dagger}(\omega)\hat{b}(\omega)\rangle_{\textrm{th}} = (e^{\hbar\beta \omega} - 1)^{-1} = \bar{N}(\omega)\), which is the Bose-Einstein thermal occupation of photons.

For an Ohmic bath, the spectral density is given with a cut-off for large values of $\omega$ by       
\[
S(\omega) = g^2(\omega) = \eta \omega e^{-\omega/\omega_c}.
\]
The real part of the spectral function $\gamma(\omega)$ is obtained by Fourier-transforming the correlation function and integrating over all modes      
\begin{eqnarray*}
    \gamma(\omega) 
    &=& \int_{-\infty}^{\infty}dt\, e^{i \omega t} \iint_0^{\infty} d\omega' d\omega'' \, C_{\omega'\omega''}(t) \\
    &=& \int_0^{\infty}d\omega' \, S(\omega')\left[ 2\pi \delta(\omega - \omega') (1 + \bar{N}(\omega'))\right. \\
    &\qquad& \qquad \qquad \qquad \qquad \left.2\pi \delta(\omega + \omega') \bar{N}(\omega') \right]\\
    &=& \begin{cases}
        2\pi \eta \omega e^{-\omega/\omega_c} (1 + \bar{N}(\omega)) & \text{for } \omega > 0 \\
        2\pi \eta |\omega| e^{-|\omega|/\omega_c} \bar{N}(|\omega|) & \text{for } \omega < 0. 
    \end{cases}
\end{eqnarray*}
We can write conditions for both positive and negative $\omega$ as      
\begin{equation}
    \gamma(\omega) 
    = \frac{2\pi \eta \omega e^{-|\omega|/\omega_c}}{1 - e^{-\beta \hbar \omega}} \label{eq:gamma_omega}
\end{equation}
which satisfies the Kubo-Martin-Schwinger (KMS) relation      
\[
\gamma(-\omega) = e^{-\beta \hbar\omega} \gamma(\omega).
\]

\setcounter{figure}{0}  
\renewcommand{\thefigure}{C\arabic{figure}}
\section{The energy, entropy, and ergotropy of the load following classical biased diffusion}
\label{app: classical biased diffusion analysis}
\subsection{Partial differential equation for classical biased diffusion}

We define a probability distribution $p(n, t)$ that tells us the probability of being in the $n$th energy level of an oscillator that changes with time $t$. To find the equation of motion for this distribution, we consider a specific level $n$ at time $t$, and consider all the processes that cause it to change. We can thus evaluate the rate of change of the probability of this level at time $t$ as the sum of all incoming transitions from the other levels minus the sum of all outgoing transitions from the current level, which gives 
\begin{equation}
    \frac{\partial p(n, t)}{\partial t} = \sum_{n'} \left[ W(n|n') p(n', t) - W(n'|n) p(n, t) \right],
    \label{eq: markov master equation}
\end{equation}
where \( W(n|n') \) is the transition rate from state \( n' \) to state \( n \). For a biased diffusion process, we consider nearest-neighbor transitions \( n \rightarrow n+1 \) with rate \( r_+ \) ,and \( n \rightarrow  n-1 \) with rate \( r_- \).
The evolution of the distribution then becomes 
\begin{equation}
    \frac{\partial p(n, t)}{\partial t} = r_- p(n-1, t) + r_+ p(n+1, t) - (r_+ + r_-) p(n, t).
\end{equation}
Treating \( n \) as a continuous variable, we expand \( p(n \pm 1, t) \) using a Taylor series        
\begin{equation}
    p(n \pm 1, t) \approx p(n, t) \pm \frac{\partial p(n, t)}{\partial n} + \frac{1}{2} \frac{\partial^2 p(n, t)}{\partial n^2} 
    \label{eq: taylor expansion of p(n, t)}
\end{equation}
which then turns Eq.~(\ref{eq: markov master equation}) to 
\begin{equation}
    \frac{\partial p(n, t)}{\partial t} = (r_+ - r_-) \frac{\partial p(n, t)}{\partial n} + \frac{1}{2} (r_+ + r_-) \frac{\partial^2 p(n, t)}{\partial n^2}.
\end{equation}
We can then define the drift velocity as \( v = r_+ - r_- \), and the diffusion constant \( D = \frac{1}{2} (r_+ + r_-) \). Thus, the partial differential equation becomes 
\begin{equation}
    \frac{\partial p(n, t)}{\partial t} = D \frac{\partial^2 p(n, t)}{\partial n^2} - v \frac{\partial p(n, t)}{\partial n}.
    \label{eq: classical biased diffusion}
\end{equation}
\subsection{Analytical Expressions for Energy, Entropy, and Ergotropy from Classical Biased Diffusion}

The solution to the biased diffusion equation in Eq.~(\ref{eq: classical biased diffusion}), assuming an initial probability distribution localized at level $n_0$ at time $t = 0$, is given by
\begin{equation}
    p(n, t) = \frac{1}{\sqrt{4\pi D t}} \exp\left(-\frac{(n - vt - n_0)^2}{4Dt}\right).
    \label{eq: probability distribution of load}
\end{equation}
This represents a Gaussian distribution centered at $n_0 + vt$, with a variance that grows linearly in time.

The average energy of the load, assuming harmonic oscillator energy levels $E_n = \hbar\omega_l (n + \tfrac{1}{2})$, is defined as
\begin{equation}
    E(t) = \sum_{n=0}^\infty \hbar\omega_l \left(n + \frac{1}{2} \right) p(n, t).
\end{equation}
When the probability distribution $p(n, t)$ varies slowly over the scale of $n$, the sum can be well approximated by a continuous integral:
\begin{equation}
    E(t) \approx \int_0^\infty \dd{n} \, \hbar\omega_l \left(n + \frac{1}{2} \right) p(n, t).
\end{equation}
The integral form is useful for deriving analytical expressions in the long-time or broad-distribution limit.
Substituting \( p(n, t) \) and letting \( \mu(t) = vt + n_0 \), \( \sigma(t) = \sqrt{2 D t} \), we can write
\begin{eqnarray}
    E_l(t) &=& \frac{\hbar\omega_l}{\sqrt{\pi D t}} \left[ \int_0^\infty n e^{-(n - \mu(t))^2 / 4 D t} \dd{n} \right.\nonumber\\
    &\qquad& \qquad \quad + \frac{1}{2} \left.\int_0^\infty e^{-(n - \mu(t))^2 / 4 D t} \dd{n} \right].
\end{eqnarray}
We see that the relevant Gaussian integrals for the above expressions are
\begin{align}
    \int_0^\infty &n e^{-(n - \mu(t))^2 / 4 D t} \dd{n} = \nonumber\\
    &\sqrt{4\pi D t} \, \mu(t) \left( 1 + \erf\left(\frac{\mu(t)}{2 \sqrt{D t}}\right)\right) + 2 D t e^{-\mu(t)^2 / 4 D t}, 
\end{align} 
and 
\begin{align}     
    \int_0^\infty & e^{-(n - \mu(t))^2 / 4 D t} \dd{n} = \sqrt{4\pi D t} \left(1 + \erf\left(\frac{\mu(t)}{2 \sqrt{D t}}\right)\right).
\end{align}
Considering a distribution that is away from the boundary $n=0$, we can safely discard the tail terms $\erf(\cdot)$ and $e^{-\mu(t)^2}$ that arise from this integral. The energy thus simplifies to 
\begin{equation}
    E_l(t) = \hbar\omega_l \left[ \mu(t) + \frac{1}{2} \right].
\end{equation}

The Von-Neumann entropy for our system that is diagonal in the energy basis $n$ is approximated in the continuum limit as 
\begin{equation}
    S(t) = -\int_0^\infty \dd{n} \, p(n, t) \ln p(n, t).
    \label{eq: continuum vn entropy}
\end{equation}
\begin{figure}[htp]
    \centering
    \includegraphics[width=0.4\textwidth]{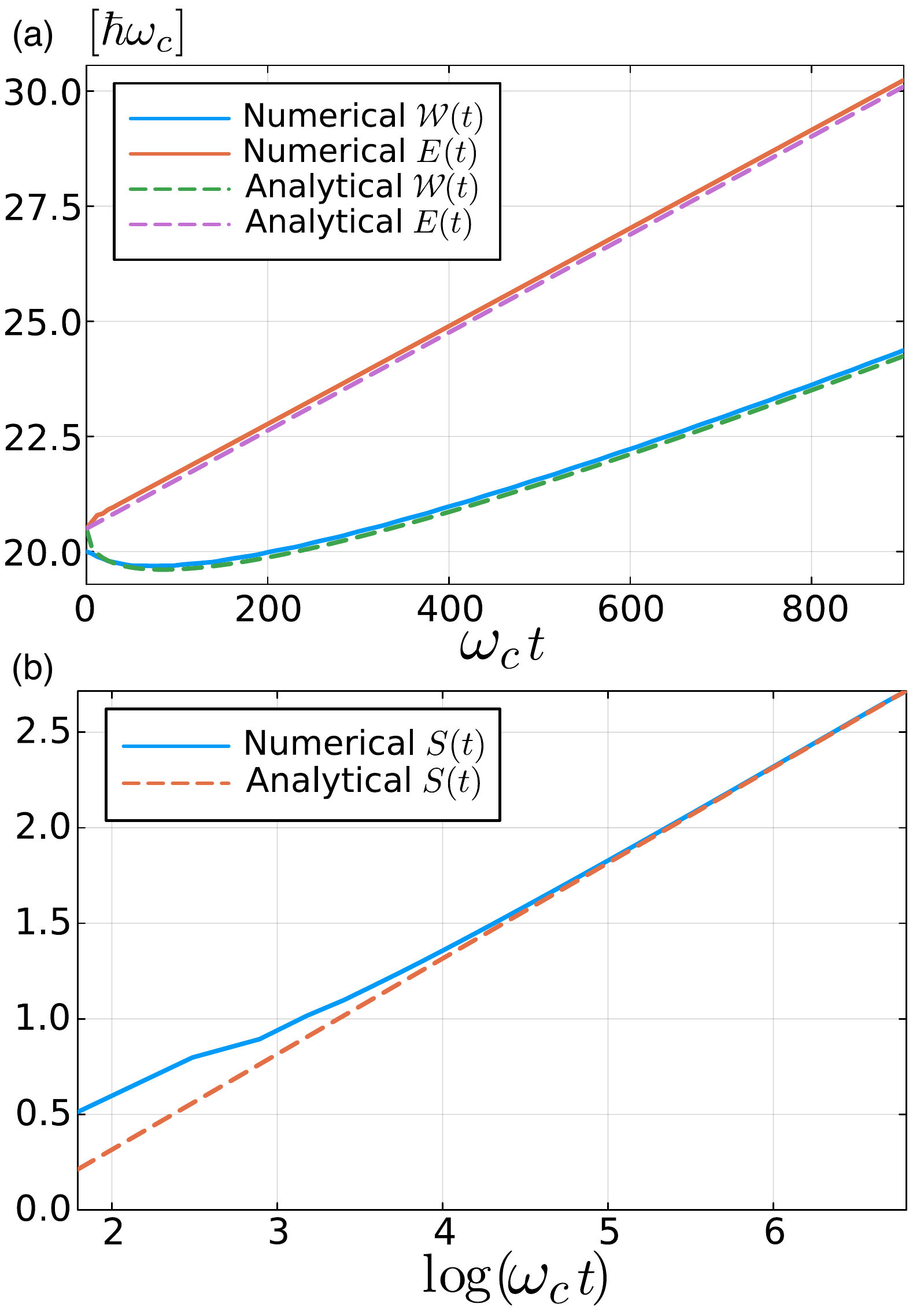}
    \caption{Comparison of the theoretical predictions with the numerical values for load's ergotropy $\mathcal{W}$, energy $E$, and entropy $S$. 
    Parameters used are: $\beta_h = 0.2(\hbar\omega_c)^{-1}, \;\beta_c = 2.0(\hbar\omega_c)^{-1}, \;\omega_l =\omega_e = \omega_c, \gamma = 0.05\omega_c$. 
    }
    \label{fig: local-lindblad-dynamics-agrees}
\end{figure}
Substituting \( p(n, t) \) from Eq.~(\ref{eq: probability distribution of load}) in Eq.~(\ref{eq: continuum vn entropy}), we get 
\begin{equation}
    S(t) = \frac{1}{2}\left[ 1 + \ln (4\pi D t)\right].
\end{equation}
To arrive at the passive state \( \pi(n, t) \), we remove the off-shift $n_0$ and center the Gaussian around zero. We then reorder the Gaussian curve to pair the negative $n$ with the positive $n$. This leads to a half-Gaussian curve with double the variance 
\begin{equation}
    \pi(n, t) = \frac{1}{\sqrt{ 2 \pi D t}} e^{-n^2 / 8 D t}.
\end{equation}
The passive energy \( E_{\pi}(t) \) is 
\begin{equation}
    E_{\pi}(t) = \int_0^\infty \dd{n} \, \hbar\omega_l \left(n + \frac{1}{2}\right) \pi(n, t).
\end{equation}
Evaluating the integral, we get  
\begin{align}
    E_{\pi}(t) &= \hbar\omega_l \left[ \sqrt{8 D t / \pi} + \frac{1}{2} \right], 
\end{align}
and, thus, the ergotropy \( \mathcal{W}_l(t) \) is 
\begin{equation}
    \mathcal{W}_l(t) = \hbar\omega_l \left( \mu(t) - \sqrt{\frac{8 D t}{\pi}} \right).
\end{equation}
These analytic results are in excellent agreement with the numerical results, besides an overall shift by a constant, see Fig.~\ref{fig: local-lindblad-dynamics-agrees}(a,b) respectively for energy/ergotropy and entropy. This constant shift in energy comes from the transient state where the load's energy might deviate from the value predicted by the biased diffusion and is carried on for later times. 

\setcounter{figure}{0}  
\renewcommand{\thefigure}{D\arabic{figure}} 
\section{The behavior of the autonomous thermal machine under weak coupling to the load}\label{app:lme}

In the weak coupling limit between the machine and the load, we can write a local master equation in GKSL form. In this regime, whereby $g \ll \omega_e,\omega_l$, we can remove the counter-propagating term and write the Hamiltonian of the machine and load as 
\begin{align}
    \hat{H} =& \underbrace{\sum_{i=1}^3 E_i \ketbra{i}{i}}_{\hat{H}_M} + \underbrace{\hbar\omega_l \left( \hat{a}^{\dagger}\hat{a} +\frac{1}{2}\right)}_{\hat{H}_L} \nonumber\\ 
    &+ \underbrace{g(\ketbra{2}{3} \otimes \hat{a} + \ketbra{3}{2} \otimes \hat{a}^{\dagger})}_{\hat{H}_{I}}. 
\end{align}
The dynamics would then be described by the local master equation 
\begin{equation}
    \dot{\hat{\rho}}(t) = \frac{i}{\hbar}[\hat{\rho}(t), \hat{H}] + \sum_{j=h,c} \mathcal{L}(\hat{A}_j),
    \label{eq: local lindblad three-level}
\end{equation}
with jump operators $\hat{A}_h = \ketbra{1}{3} \otimes \hat{I}_l$ (for the hot bath) and $ \hat{A}_c = \ketbra{1}{2} \otimes \hat{I}_l$ (for the cold bath). The dissipators for bosonic thermal baths are  
\begin{equation}
    \mathcal{L}
    (\hat{A}_j) = \gamma \left[ \bar{N}_i(\omega) \Gamma[\hat{A}_i] \\
        + (1+\bar{N}_i(\omega)) \Gamma[\hat{A}_i^\dagger] \right],
\end{equation}
where $\Gamma[\hat{O}] \equiv \hat{O}\hat{\rho}\hat{O}^\dagger - \tfrac{1}{2}\{\hat{O}^\dagger\hat{O}, \hat{\rho}\}$ and $\bar{N}_i(\omega)$ is the thermal average occupation of the oscillator with frequency $\omega$ at inverse temperature $\beta_i$. 

The jump rates $\gamma$ for transitions at Bohr frequency $\omega$ are given by Eq.~(\ref{eq:gamma_omega}), which in this context become the rates for upward and downward transition rates $\gamma_{\alpha}(\omega)$ for Lindblad operators $\hat{\sigma}_\pm$ for the levels $i$ and $j$ and can be written as       
\begin{eqnarray}
\gamma_{+}(\omega) &=& \gamma_0\, \bar{N}(\omega), \nonumber \\
\gamma_{-}(\omega) &=& \gamma_0\, (1 + \bar{N}(\omega)). 
\label{eq: local-jump-coefficients}
\end{eqnarray}

The steady state of the three-level system without finite-size effects has a non-thermal population, while the load performs a biased diffusion with a linear change of the mean and variance in its energy occupations, as observed in Refs.~\cite {scarani2017, colinTeo2017, QuantumLoad2020}. We also observe that in this framework, the reduced density matrices of both the load and the engine are diagonal in their computational bases, despite the total density matrix having coherent terms in the off-diagonals. This allows us to perform further analytic calculations by considering the load's energy distribution as a solution to a classical biased diffusion equation described by Eq.~(\ref{eq: classical biased diffusion}).
\begin{figure}[htp]
    \includegraphics[width=0.8\linewidth]{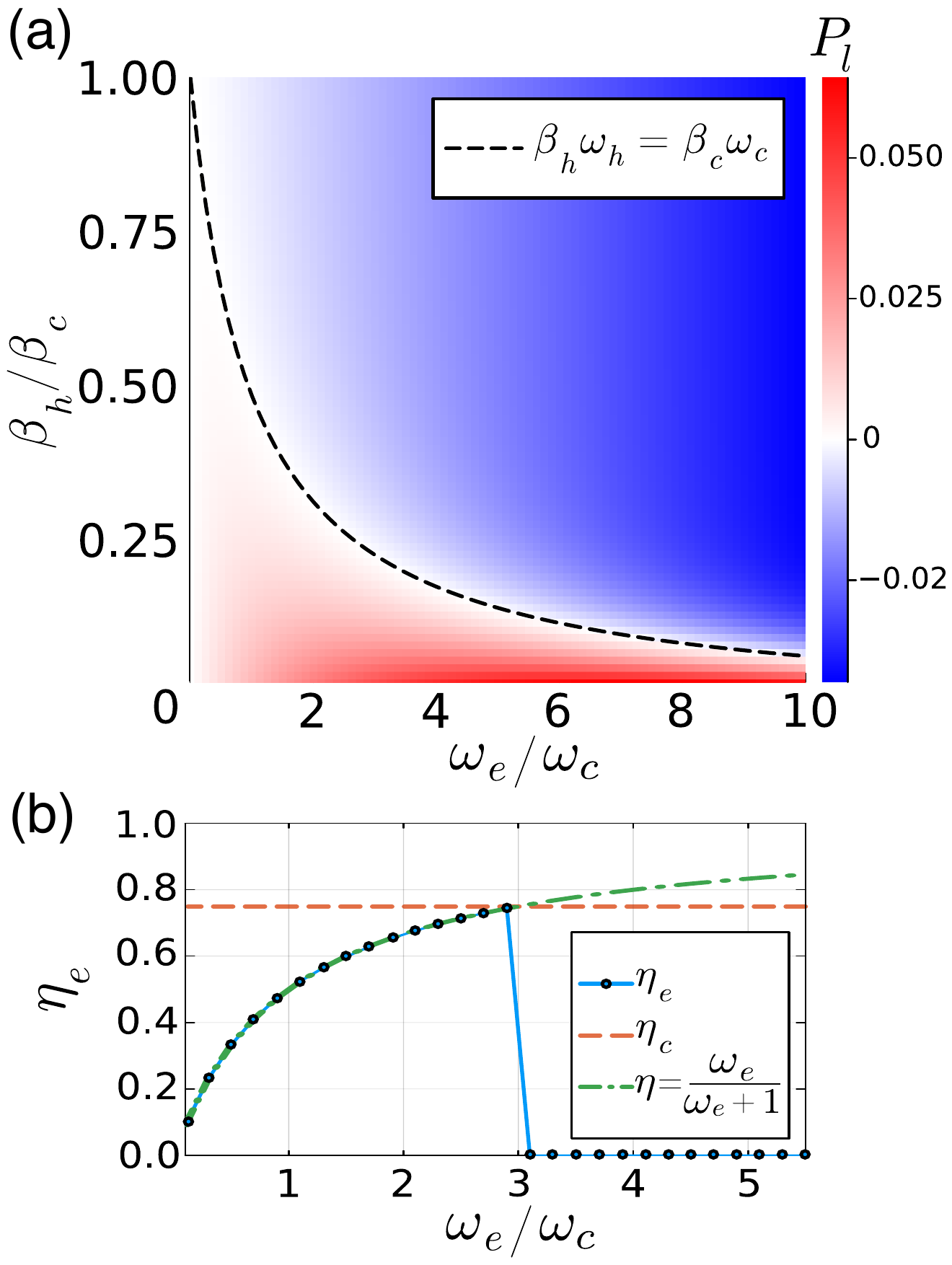}
    \label{fig:phasespace}
    \caption{
    (a) Heatmap of the load's rate of energy increase $P_l$ as a function of $\beta_h$ and $\omega_e$, with red (positive power) and blue (negative power) regions indicating engine and refrigerator operation, respectively. White denotes near-zero power. The theoretical transition curve given by Eq.~(\ref{eq:eng_ref_condition}) is overlaid. 
    (b) Efficiency of the thermal machine when functioning as an engine versus $\omega_e$ (solid line), while the green dot-dashed line represents a mathematical continuation of the curve given by Eq.~(\ref{eq: engine efficiency}), and the red dashed line represents Carnot efficiency. In (b), we considered a cut along the horizontal line $\beta_h/\beta_c = 1/4$. Common parameters are $\beta_c=4.0(\hbar\omega_c)^{-1}, \;g = 0.05\hbar\omega_c, \gamma = 0.05\omega_c$. 
    }
    \label{fig: Phase space and carnot efficiency}
\end{figure}

We systematically explore the engine's phase space to identify operational regimes (engine vs. refrigerator), as depicted in Fig.~\ref{fig: Phase space and carnot efficiency}(a). Scovil and Schulz~\cite{Scovil1959} have derived the expression that tells us when a three-level maser system coupled to two baths will change from an engine to a refrigerator by evaluating the ratio of the particle population between the second and the third level. Assuming the three levels are in thermal equilibrium, one would get
\begin{align*}
     \frac{n_3}{n_2} & = \frac{n_3}{n_1}\frac{n_1}{n_2}
     = e^{\beta_c\omega_c-\beta_h\omega_h}.
\end{align*}
Population inversion ($n_3/n_1 > 1$) occurs when $\beta_h\omega_h < \beta_c\omega_c$, marking the engine regime where positive work is performed. Conversely, when $\beta_h\omega_h > \beta_c\omega_c$ ($n_3/n_1 < 1$), the system loses the population inversion and acts as a refrigerator. 
This results in the boundary between the engine and refrigerator functioning modes given by 
\begin{equation}
    \beta_h\omega_h = \beta_c\omega_c. \label{eq:eng_ref_condition}
\end{equation}
Similarly, one can evaluate the efficiency of the engine functioning regime from 
\begin{equation}
    \eta_e = \frac{P_l}{\dot{Q}_h} = \frac{\omega_e}{\omega_h} = 1 - \frac{\omega_c}{\omega_h}.
    \label{eq: engine efficiency}
\end{equation}
When $\eta_e$ exceeds the Carnot efficiency $\eta_c = 1 - \beta_h/\beta_c$, the system switches to refrigeration, which matches Eq.~(\ref{eq:eng_ref_condition}). This transition is evident in Fig.~\ref{fig: Phase space and carnot efficiency}(b). For smaller $\omega_e$, when $\eta_e < \eta_c$, we observe finite efficiency given by Eq.~(\ref{eq: engine efficiency}). As $\omega_e$ reaches the value such that $\eta_e=\eta_c$, the energy flows change, and the efficiency drops to 0. 
We must note that despite the three-level system reaching the Carnot efficiency, the heat currents and the power pumped into the load are very small in these regions. 

\begin{figure}[hp]
    \centering
    \includegraphics[width=\linewidth]{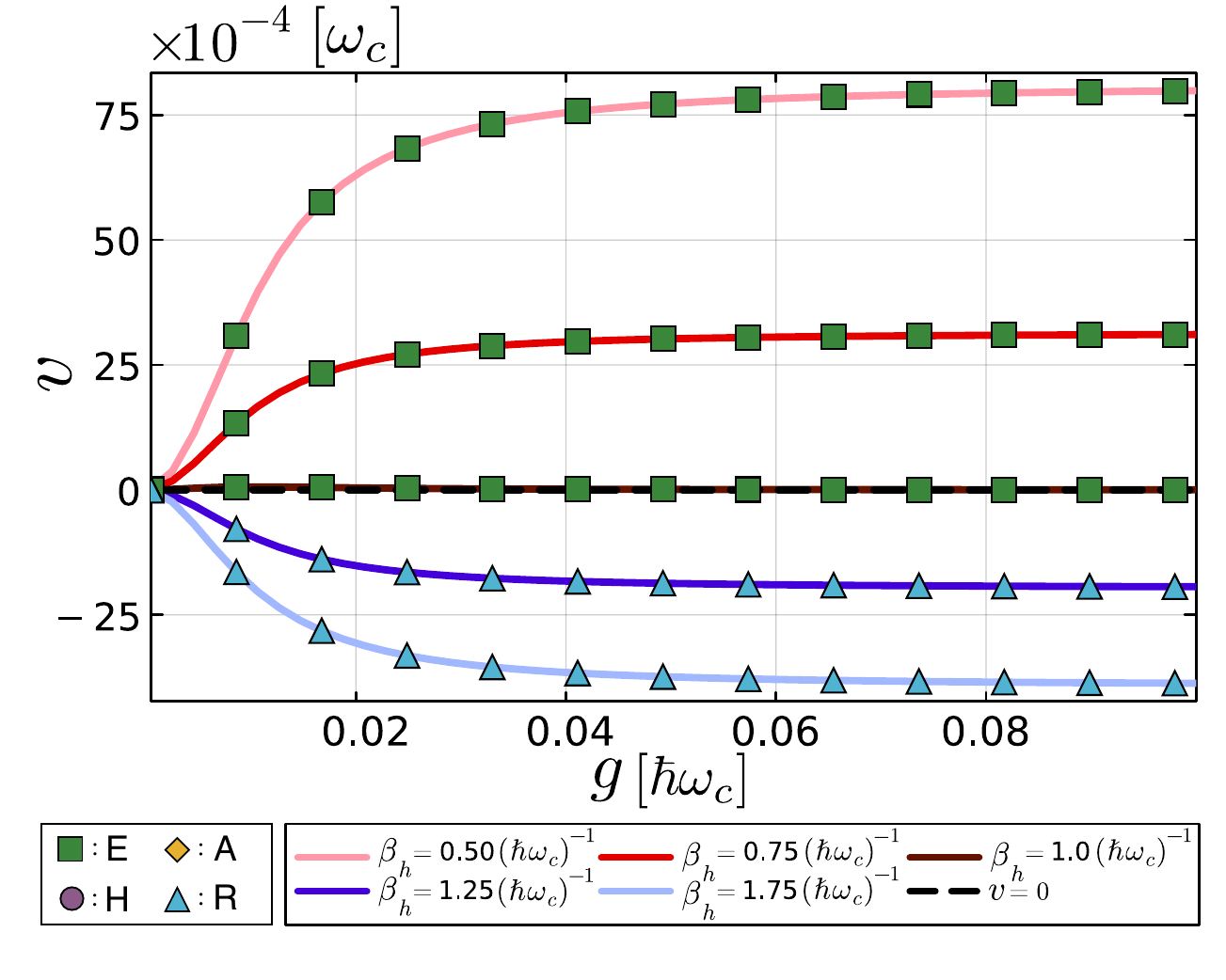}
    \caption{The effect of coupling strength $g$ on the drift velocity $v$ of the load plotted for various value of $\beta_h$ for $\beta_c = 2.0 (\hbar\omega_c)^{-1}, \omega_l = \omega_e = \omega_c, \gamma = 0.05\omega_c$. 
    }
    \label{fig: local-lindblad-coupling-effect}
\end{figure}

We now focus on the effect of coupling strength on the functioning of the thermal machine, in the weak machine-load coupling regime. Specifically, we study the dependence of $v$ on $g$. The result is shown in Fig.~\ref{fig: local-lindblad-coupling-effect}. The power (which is proportional to $v$) increases quadratically for $g\ll 1$, and then reaches a plateau and saturates to a constant value. The value of $v$ is always positive for the engine functioning regime, and negative for the refrigerator one. 
This informs us that in a weak-coupling description of the three-level system, the coupling strength does not change the functioning of the three-level system from engine to refrigerator and vice-versa, nor allows the emergence of the accelerator and heater functioning regimes.

\end{appendix}

\end{document}